\documentclass[aps,prl,reprint,amsmath,amssymb,ap,superscriptaddress]{revtex4-2}

\usepackage[T1]{fontenc}
\usepackage{gibbon}
\usepackage{xr}
\usepackage{graphicx}
\usepackage{float}
\usepackage{color}
\usepackage{xr}

\begin{document}

	\title{Irreversiblity in Bacterial Turbulence: Insights from the Mean-Bacterial-Velocity Model}
	
	\author{Kolluru Venkata Kiran}
	\email[]{kollurukiran@iisc.ac.in}
		\affiliation{Centre for Condensed Matter Theory, Department of Physics, Indian Institute of Science, Bangalore, 560012, India. }
	
	\author{Anupam Gupta}
	 \email[]{agupta@phy.iith.ac.in}
	\affiliation{Department of Physics,
Indian Institute of Technology (IIT), Hyderabad,
Kandi Sangareddy
Telangana, 502285, India}
   
	\author{Akhilesh Kumar Verma}
	\email[]{akvermajnusps@gmail.com}
	\affiliation{Mathematics Institute, Zeeman Building, University of Warwick, Coventry CV4 7AL, UK}
	
	\author{Rahul Pandit}
		\email[]{rahul@iisc.ac.in}
	\affiliation{Centre for Condensed Matter Theory,Department of Physics, Indian Institute of Science, Bangalore, 560012, India. }

	\begin{abstract}
		We use the mean-bacterial-velocity model to investigate the \textit{irreversibility} of two-dimensional (2D) \textit{bacterial turbulence} and to compare it with its 2D fluid-turbulence counterpart. We carry out extensive direct numerical simulations of Lagrangian tracer particles that are advected by the velocity field in this model. Our work uncovers an important, qualitative way in which irreversibility in bacterial turbulence is different from its fluid-turbulence counterpart: For large positive (or large but negative) values of the \textit{friction} (or \textit{activity}) parameter, the probability distribution 
		functions of energy increments, along tracer trajectories, or the power are \textit{positively} skewed; so irreversibility in bacterial turbulence can lead, on average, to \textit{particles gaining energy faster than they lose it}, which is the exact opposite of what is observed for tracers in 2D fluid turbulence. 
	\end{abstract}

	\maketitle

	\section{}

	Most fluid flows are turbulent; and they can attain a nonequilibrium, but statistically steady, state (NESS), if the energy injection into the fluid, say by an external force, is balanced by viscous dissipation. Far away from boundaries, this NESS is statistically homogeneous and isotropic if we consider length scales that are much smaller than the energy-injection scale $l_f$\cite{frisch1995turbulence,rose1978fully}. Two important characteristics of this NESS are (a) the distribution of energy over a large range of length scales and (b) the temporal irreversibility of turbulent flows. This irreversibility is not easily apparent if we look at movies, played forward or backward in time, of Lagrangian particles, or \textit{tracers}, that
	are advected by turbulent flows; however, the statistics of such tracers or inertial particles in turbulent flows yields signatures of this irreversibility~\cite{xu2014flight,jucha2014time,chertkov1999lagrangian,xu2016lagrangian,pumir2016single,falkovich2013single,bhatnagar2018heavy,pietrzyk2022analysis}:
	if we analyse (a) the increments 
	\begin{equation}
		W(t,\tau)\equiv E(t+\tau)-E(t)    
		\label{eq:W}
	\end{equation}
	of the particle energy $E$ at time $t$ or (b) the power 
	\begin{equation}
		p_{L}(t)\equiv\frac{d E}{dt} = a_L v_L,
		\label{eq:pL}
	\end{equation}
	with $v_L$ the magnitude of the tracer velocity and $a_L$ the component of its acceleration along its trajectory. It has been found that probability distribution functions (PDFs) of $W$ and $p_L$, obtained by averaging over $t$ and the trajectories of all tracers, are negatively skewed~\cite{pumir2016single,xu2014flight,bhatnagar2018heavy,picardo2020lagrangian,ray2018non,vsvanvcara2019flight,verma2021heavy}; i.e., on average, such tracers lose energy faster than they gain it. Is it possible to use these PDFs to characterise irreversibility in \textit{bacterial turbulence}~\cite{Wensink14308,Dunkel_2013,dunkel2013fluid,slomka2015generalized,linkmann2019phase,linkmann2020condensate,slomka2017spontaneous,bratanov2015new}? We show that this can, indeed, be done. We illustrate this by carrying out an extensive study of the irreversibility of bacterial turbulence in the mean-bacterial-velocity model~\cite{Wensink14308}. Our work uncovers an important, qualitative way in which irreversibility in bacterial turbulence is different from its fluid-turbulence counterpart: For large positive (or large but negative) values of the \textit{friction} (or \textit{activity}) parameter $\alpha$ (see below), the PDFs of $W(\tau)$ or $p_L$ are \textit{positively} skewed; this implies that irreversibility in bacterial turbulence can lead, on average, to \textit{particles gaining energy faster than they lose it}, for certain ranges of values of $\alpha$. 

	Dense bacterial suspensions, which are examples of active systems~\cite{Wensink14308,Dunkel_2013,linkmann2020condensate,oza2016generalized,rana2020coarsening,alert2021active}, show spatiotemporal evolution that is reminiscent of flows in turbulent fluids. Hydrodynamical models have been developed to describe turbulence in dense, quasi-two-dimensional (2D) bacterial suspensions~\cite{dunkel2013fluid,slomka2015generalized,linkmann2019phase,linkmann2020condensate,slomka2017spontaneous,rana2020coarsening,thampi2013velocity,thampi2016active}.
	We use the mean-bacterial-velocity model~\cite{Wensink14308} or the Toner-Tu-Swift-Hohenberg (TTSH) model~\cite{alert2021active,bar2020self}, for the incompressible velocity field $\bu(\mathbf{x},t)$; this model has been employed to study turbulence in dense suspensions of \textit{B. subtilis}:
	\begin{eqnarray}\label{eq:maineq}
		\frac{\partial\bu}{\partial t}+\lambda_{0}\bu.\mathbf{\nabla}\bu &=&-\nabla P-(\alpha+\beta|u|^{2})\bu \nonumber \\
		&+&\Gamma_{0}\nabla^{2}\bu-\Gamma_{2}\nabla^{4}\bu; \nonumber \\
		\nabla \cdot \bu &=& 0.
		\label{eq:TTSH}
	\end{eqnarray}
	Here,  $P(\mathbf{x},t)$ is the pressure at point $\mathbf{x}$ and time $t$; the constant density $\rho$ is set to unity~\footnote{ Equation (\ref{eq:maineq}) is not Galilean invariant; it 
		reduces to the Navier-Stokes equation with friction for  
		$\Gamma_{0}>0, \alpha>0, \Gamma_{2}=0, \lambda_{0}=1$, and $\beta=0$.}. We use periodic boundary conditions in all directions because we concentrate on statistically homogeneous and isotropic bacterial turbulence. We restrict ourselves to two dimensions (2D) as most experiments in this field have been conducted in quasi-2D systems.

	The parameters $\Gamma_{0}<0$ and $\Gamma_{2}<0$; a spatial Fourier transform of Eq.~(\ref{eq:TTSH}), followed by a linear-stability analysis about the spatially uniform state, yields the wave vectors $\mathbf{k}$, with magnitude $k$, for which there are linearly unstable modes. We define the following
	characteristic length, velocity, and time scales, respectively:
	\begin{equation}
		\Lambda=2\pi\sqrt{\frac{2\Gamma_{2}}{\Gamma_{0}}}; \, v_{0}=\sqrt{\frac{|\Gamma_{0}|^{3}}{\Gamma_{2}}}; \,  \theta =\frac{\Lambda}{v_{0}}.
		\label{eq:scales}
	\end{equation}
	These unstable modes provide a source of energy injection into the system ~\footnote{This is similar to energy injection in the Kuramoto-Sivashinsky equation see, e.g., Refs.~\cite{kuramoto1976persistent,sivashinsky1977nonlinear,roy2020one}.}; this energy is dissipated by (a) the linearly stable modes, (b) the cubic term with the coefficient $\beta > 0$, and (c) the linear term with the coefficient $\alpha$, if $\alpha > 0$. Moreover,  there is energy injection, or \textit{activity}, if $\alpha < 0$; and $\Gamma_0 < 0$ and $\lambda_0 \neq 1$ also induce activity~\cite{bar2020self} [$\lambda_{0}>1$ for \textit{pusher swimmers} like \textit{B. subtilis} (see, e.g., Refs.~\cite{Wensink14308,Dunkel_2013,james2018turbulence})]. The interplay between these energy-injection and dissipation terms leads to a NESS with self-sustained, turbulence-type patterns~\cite{bratanov2015new}. The effective viscosity
	\begin{equation}{\label{eq:effvis}}
		k^2 \nu_{eff}(k)=\big(\alpha+2\beta u^{2}_{rms}+\Gamma_{0} k^2 +\Gamma_{2}k^4\big)
	\end{equation}
	can be used to rewrite Eq.~(\ref{eq:TTSH}) in a Navier-Stokes form (see the Supplemental Material~\ref{SM} and Ref.~\cite{bratanov2015new}).
	Clearly, the wave numbers $k$ at which energy is injected
	(dissipated) are those with $\nu_{eff}(k)<0 ~(>0)$; the root-mean-square velocity $u_{rms}$
	must be obtained from a calculation (see below).
	
	We solve Eq.~(\ref{eq:TTSH}) by a pseudospectral direct numerical simulation (DNS) [see, e.g., Ref.~\cite{perlekar2011persistence}] with $N^2=1024^2$ collocation points and the parameters in Table~\ref{tab:parameters}; we have checked in representative cases that our results are unchanged if we use $N^2 = 2048^2$ collocation points. We hold $\lambda_0$, $\beta$, and $\Gamma_0$ fixed, and we tune the activity principally by varying $\alpha$.
	
	\begin{table}
		\centering
		\begin{tabular}{|c| c| c| c|c|c| c|}
			\hline
			Run & $\Gamma_{2}$ &  $\alpha$& $\delta t$& $\Lambda$ & $\theta$ & $v_{0}$
			\\
			\hline
			A1-A12& 9e-5& $\ast$&2e-4&0.40&0.40&1.0\\
			\hline
			B& 9e-5&  -4&1e-5&0.40&0.40&1.0\\
			\hline
			C& 9e-5&  1&5e-5&0.40&0.40&1.0\\
			\hline
			D& 3.6e-5& 14&1e-4&0.25&0.16&1.60\\
			\hline
		\end{tabular}
		\caption{\label{tab:widgets} Parameters for our DNSs:$^\ast$~Runs A1-A12 use $\alpha=-4,\,-3.5,\,-2.5,\,-1.5,\,-1,\,-0.5,\,0,\,1,\,3,\,3.5,\,4,$ and $5$, respectively. For all the Runs listed in the table, $\Gamma_{0}=-0.045$, $\beta=0.5$, and $\lambda_{0}=3.5$.}
		\label{tab:parameters}
	\end{table}
	
	\begin{figure*}[!]	
	 
		\resizebox{\linewidth}{!}{			{
	\includegraphics[scale=0.4]{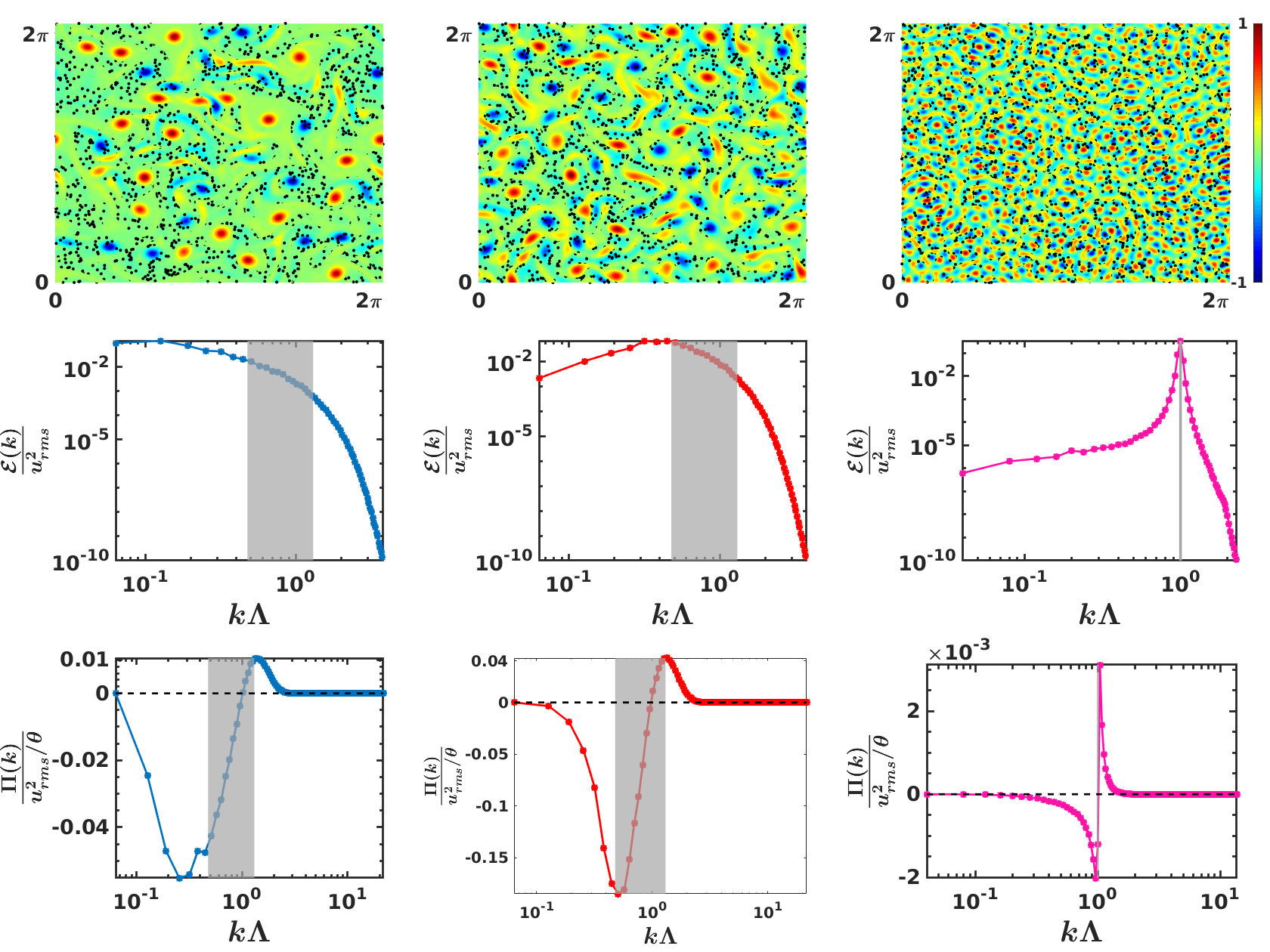}
		\put(-480,375){\rm {\bf(a)}}
		\put(-310,375){\rm {\bf(b)}}
		\put(-140,375){\rm {\bf(c)}}
		\put(-480,240){\rm {\bf(d)}}
		\put(-310,240){\rm {\bf(e)}}
		\put(-140,240){\rm {\bf(f)}}
		\put(-482,125){\rm {\bf(g)}}
		\put(-312,125){\rm {\bf(h)}}
		\put(-142,125){\rm {\bf(i)}}
}
}
	\caption{(Color online) Plots for runs A1, A8, and D (Table~\ref{tab:parameters}):  (a)-(c) Filled contour plots of  the vorticity $\bom(\mathbf{x},t))$, with some tracers (black points), at a representative time in the NESS; log-log (base 10)  plots versus $k\Lambda$ of (d)-(f) the energy spectrum $\mathcal{E}(k)$ and (g)-(i) the energy flux $\Pi(k)$ [Eq.~(\ref{eq:spect})]; the gray-shaded areas indicate the ranges of $k$ for which 
		$\nu_{eff}(k)<0$ [Eq.~(\ref{eq:effvis})].}
	\label{fig:ene_spect}
	
	\end{figure*}
	\begin{figure}[!]	
		\resizebox{\linewidth}{!}{
	\includegraphics[scale=0.8]{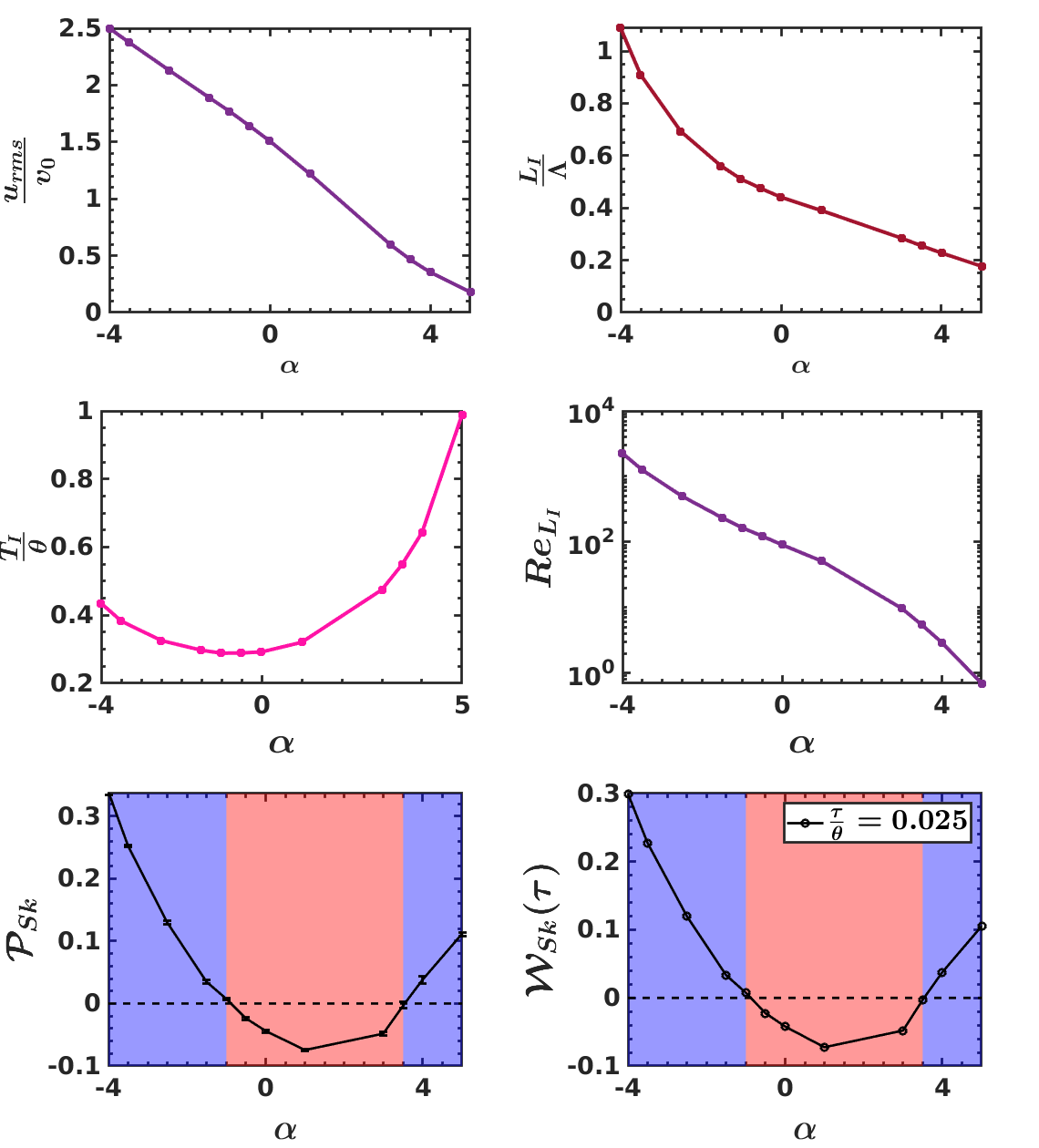}
\put(-180,225){\rm {\bf(a)}}
\put(-75,225){\rm {\bf(b)}}
\put(-180,140){\rm {\bf(c)}}	
\put(-75,140){\rm {\bf(d)}}
\put(-180,60){\rm {\bf(e)}}
\put(-75,60){\rm {\bf(f)}}

}
		\caption{(Color online) Plots versus $\alpha$ of (a) $u_{rms}/v_{0}$, (b) $L_{I}/\Lambda$, (c) $T_I/\theta$, (d) $Re_{L_I}$, (e) the skewness $\mathcal{P}_{Sk}$, and (f) 
			$\mathcal{W}_{Sk}(\tau)$ for $\tau/\theta=0.025$; in (e) and (f), blue and pink shading indicate, respectively, ranges of $\alpha$ in which the skewnesses are  positive and negative.}
		
		\label{fig:Pskn_pdf}
	\end{figure}
	\begin{figure*}[!]	
		
		\resizebox{\linewidth}{!}{
				\includegraphics[scale=0.4]{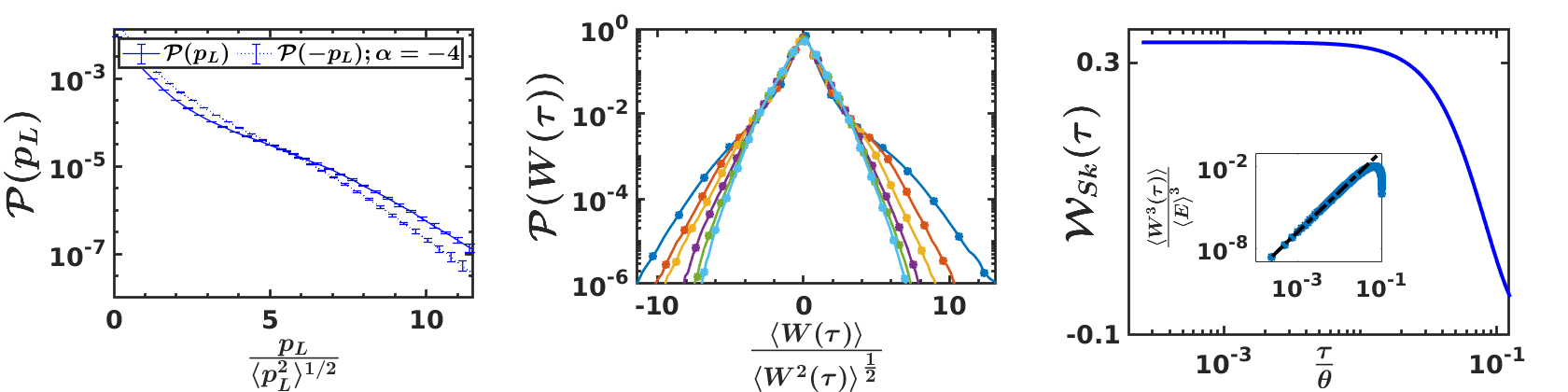}
					\put(-480,120){\rm {\bf(a)}}
					\put(-300,120){\rm {\bf(b)}}
					\put(-150,120){\rm {\bf(c)}}
		}
		\caption{(Color online) Plots for run B: (a) Semi-log plot of the normalized PDFs (a) $\mathcal{P}(p_{L})$ and (b) $\mathcal{P}(W(\tau))$, with $\tau/\theta$ going from $0.025, 0.08, 0.13, 0.25, 0.38,$ to $0.50$, as we move from the outermost to the innermost curve; in (a) negative values of $p_{L}$ (dashed) are reflected about the  vertical axis to highlight the asymmetry of $\mathcal{P}(p_{L})$. 
		$(c)$ Log-Log (base 10) plot versus $\tau/\theta$ of the skewness $\mathcal{W}_{Sk}(\tau)$. Inset: for the same range of $\tau/\theta$, a log-log plot versus $\tau/\theta$ of $\langle W^3(\tau)\rangle/\langle E \rangle^3$; the dashed black line is a fit to  $\langle W^3(\tau)\rangle/\langle E \rangle^3 \sim (\tau/\theta)^{3}$.}
		\label{fig:ene_inc}
	\end{figure*}

\begin{figure}[!]	
		{\includegraphics[scale=0.5]{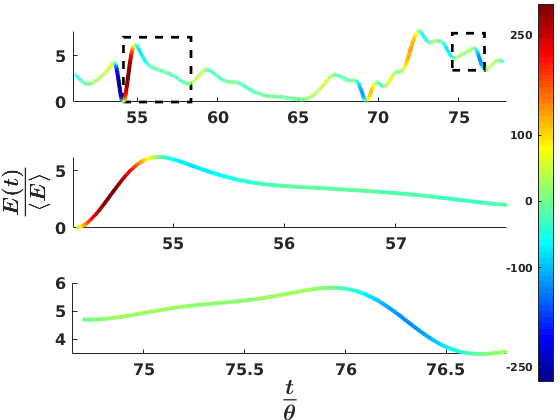}	
			\put(-183,140){\rm {\bf(a)}}
			\put(-183,90){\rm {\bf(b)}}
			\put(-183,50){\rm {\bf(c)}}
			\put(0,90){\bf {$p_{L}$}}
		}
		
		\caption{ (Color Online) Plot of the normalised energy $E(t)/\langle E \rangle$ versus
			the normalized time $t/\theta$ along a representative particle trajectory from Run A1; the colors along the trajectory indicate the value of $p_{L}$. }
		\label{fig:color}
	\end{figure}
	
	In Figs.~\ref{fig:ene_spect} (a), (b), and (c), we present filled contour plots of  the vorticity $\bom(\mathbf{x},t)=\mathbf{\nabla}\times\bu(\mathbf{x},t)$, with some tracers shown via black points, for the representative Runs A1, A8, and D, respectively (see Table~\ref{tab:parameters}); in  Figs.~\ref{fig:ene_spect} (d)-(i), we give log-log  plots versus $k\Lambda$ of the $k$-shell-averaged energy spectrum $\mathcal{E}(k)$ and energy flux $\Pi(k)$: 
	\begin{eqnarray}
		\mathcal{E}(k)&=&\frac{1}{2} \sum_{k'=k-1/2}^{k'=k+1/2}\langle \widetilde{\bu}(\mathbf{k}^{'}).\widetilde{\bu}(-\mathbf{k}^{'})\rangle_{t}; \nonumber \\
		\Pi(k)&=&-\lambda_{0}\sum_{k'=0}^{k'=k}\sum_{k''=k'-1/2}^{k''=k'+1/2} [\langle \widetilde{\bu}(-\mathbf{k}^{''})\cdot\textbf{P}(\mathbf{k''}) \cdot \nonumber \\ &(&\widetilde{\bu.\nabla\bu)}({\mathbf{k''})}\rangle_{t}];
		\label{eq:spect}
	\end{eqnarray}
	here, tildes denote spatial Fourier transforms, $\langle \cdot \rangle_t$ is the time average
	over the NESS, and the transverse projector $\textbf{P}(\mathbf{k})$ has the 
	components $P_{ij}(\mathbf{k})=\delta_{ij}-\frac{k_{i}k_{j}}{k^{2}}$. The total fluid energy, root-mean-square velocity, integral length scale, integral time scale, and integral-scale Reynolds number are, respectively,
	\begin{eqnarray}
		\mathcal{E}_T&=& \sum_k \mathcal{E}(k) ; ~ u_{rms} = \sqrt{2\mathcal{E}_T} ; 
		L_I =\frac{\sum_k [\mathcal{E}(k)/k]}{\sum_k \mathcal{E}(k)} ;\nonumber \\ 
		T_I &=& L_I/u_{rms} ; \; Re_{L_I} \equiv u_{rms} L^3_I/\Gamma_2.
		\label{eqn:Re}
	\end{eqnarray}
	The gray-shaded areas in Figs.~\ref{fig:ene_spect} (d)-(i) indicate the ranges of $k$ for which 
	$\nu_{eff}(k)<0$. For the Runs in Table~\ref{tab:parameters}, there is no range of $k$ over which $\Pi(k)$ remains constant, unlike its fluid-turbulence counterpart, so we cannot identify inverse- or forward-cascade regimes in $\mathcal{E}(k)$; however, $\mathcal{E}(k)$ is spread over a large range of $k$ and the temporal evolution of $\bu$ is chaotic, so the bacterial-turbulence NESS for this model [Eq.~(\ref{eq:TTSH})] displays spatiotemporal chaos. In Figs.~\ref{fig:Pskn_pdf} (a)-(d) we present plots  versus $\alpha$ of $u_{rms}/v_0, \, L_I/\Lambda, \, T_I/\theta,$ and $Re_{L_I}$, respectively (Runs A1-A12); as $\alpha$ moves from the \textit{activity regime} ($\alpha < 0$) to the \textit{frictional regime} ($\alpha > 0$), $u_{rms}/v_0,\, L_I/\Lambda,$ and $Re_{L_I}$ decrease, but $T_I/\theta$ first decreases and then increases because $u_{rms}$ decreases more rapidly than $L_I$.

	The velocity $\mathbf{v}_{L}(t)$ of a tracer at $\mathbf{x}_{L}(t)$ is
	\begin{equation}\label{eq:Lageq}
		\frac{d \mathbf{x}_{L}(t)}{dt}\equiv \mathbf{v}_{L}(t) = \bu(\mathbf{x}_{L}(t),t).
	\end{equation}
	We track $N_p = 10,000$ tracers, employ the second-order Runge-Kutta method
	for time marching, and evaluate $\bu(\mathbf{x}_{L}(t),t)$ at off-grid points via
	bilinear interpolation~\cite{james2018vortex, mukherjee2021anomalous,singh2021lagrangian}; to get good statistics, we use very long runs ($3\times10^6$ time steps per particle). The acceleration of a tracer particle is
	\begin{eqnarray}
		\mathbf{a}(\mathbf{x}_{L},t)&\equiv&\frac{\partial\bu}{\partial t}+(\bu.\nabla) \bu\bigg{\vert}_{x_{L}(t)} \nonumber \\ &=&-\nabla P_{eff}-(1-\lambda_{0})(\bu\times\bom) \nonumber \\ -(\alpha+\beta|u|^{2})\bu &+&\Gamma_{0}\nabla^{2}\bu-\Gamma_{2}\nabla^{4}\bu\bigg{\vert}_{x_{L}(t)}, 
		\label{eq:Lag_acc}
	\end{eqnarray}
	where the effective pressure $P_{eff}=P-\frac{1}{2}(1-\lambda_{0})\bu.\bu$; the component of this acceleration along the tracer's trajectory yields $a_L$, whence we get $p_L$ [Eq.~(\ref{eq:pL})]
	and its normalized PDF $\mathcal{P}\big(\frac{p_{L}}{\langle p^{2}_{L}\rangle^{1/2}}\big)$. From the time series of particle energies (Fig.~\ref{fig:color}) we obtain the energy-increment PDFs 
	$ \mathcal{P}\big(\frac{W(\tau)}{\langle W^{2}(\tau)\rangle^{1/2}}\big)$, for various values of $\tau<T_{I}$. Both these PDFs have zero mean (Figs.~\ref{fig:ene_inc} (a) and (b)), because we are considering a statistically steady state in which the mean energy input is balanced by dissipation, but they are asymmetrical; we characterize this asymmetry by computing the skewnesses
		\begin{equation}
		\mathcal{P}_{Sk}=\frac{\langle p^{3}_{L}\rangle}{\langle p^{2}_{L}\rangle^{\frac{3}{2}}} \;  {\rm{and}} ~~\mathcal{W}_{Sk}(\tau)=\frac{\langle W^{3}(\tau)\rangle}{\langle W^{2}(\tau)\rangle^{\frac{3}{2}}},
		\label{eq:psk}
	\end{equation}
	which we plot versus $\alpha$ in Figs.~\ref{fig:Pskn_pdf}(e) and (f), respectively. We observe that $\mathcal{P}_{Sk}>0$ in the large-activity, $\alpha<-2$, and extreme-friction, $\alpha>3.5$, regions (shaded blue). This is in \textit{stark contrast} to 2D fluid turbulence~\cite{xu2014flight} where $\mathcal{P}_{Sk}<0$. The values of $\alpha$ for which $\mathcal{P}_{Sk}<0$ lead to NESSs that are characterized by \textit{flight-crash} events in which, on average, $E(t)$ builds up slowly but decays rapidly.[In 2D Faraday-wave experiments, $\mathcal{P}_{Sk}>0$ has been attributed to the temporal coherence of these waves, and been removed by filtering \footnote{For conventional 2D fluid turbulence, Refs.~\cite{xu2014flight,pumir2014redistribution} discuss, for both DNSs and experiments, the effects of different types of forcing on $\mathcal{P}_{Sk}$; they report $\mathcal{P}_{Sk} < 0$ in DNSs with white-noise forcing; by contrast, in Faraday-wave experiments, they observe $\mathcal{P}_{Sk}>0$, which they attribute to the temporal coherence of Faraday waves. In the latter case they employ a filtering procedure that again yields $\mathcal{P}_{Sk}<0$. In the Supplemental Material~\ref{SM} we investigate the effects of a similar filtering procedure for 2D bacterial turbulence in Eq.~(\ref{eq:TTSH}).}.] For runs B and D we also find $\mathcal{P}_{Sk}>0$. In Fig.~\ref{fig:color} (a) we plot the time series of $E/\langle E \rangle$ of a typical particle. We also show magnified regions of this time series to exhibit flight-crash events [Fig.~\ref{fig:color} (c)], of the types that are predominant in fluid turbulence, and the events in which $E(t)$ builds up faster than it decays [Fig.~\ref{fig:color} (b)]. In the large-activity and extreme-friction regions mentioned above, the predominance of the events shown in Fig.~\ref{fig:color} (b) leads to $\mathcal{P}_{Sk}>0$ and, for small $\tau/\theta$, $\mathcal{W}_{Sk}(\tau)=\frac{\langle W^{3}(\tau)\rangle}{\langle W^{2}(\tau)\rangle^{\frac{3}{2}}} > 0$, because $\lim_{\tau\to0} W(\tau,t) \sim p_L(t)$. Furthermore,
	for $\tau/\theta \ll 1$, we obtain the Taylor-expansion result $\langle W^{3}(\tau)\rangle\sim \tau^{3}$, for which we give a representative plot in the inset of Fig.~\ref{fig:ene_inc} (c). As $\tau$ decreases, the tails of $ \mathcal{P}\big(\frac{W(\tau)}{\langle W^{2}(\tau)\rangle^{1/2}}\big)$ widen, as in fluid turbulence~\cite{pumir2016single}\footnote{ This widening could be a signature of intermittency effects, which we examine elsewhere~\cite{KKolluru_unpub}.}.

	The sign of $\mathcal{P}_{Sk}$ (and, for small $\tau/\theta,$ the sign of $\mathcal{W}_{Sk}(\tau)$) displays the following correlation with the scale-by-scale energy budget in Fourier space, where we can identify the $k$-dependence of the energy contributions from the terms with coefficients $\alpha$, $\Gamma_0$, and $\lambda_0$ in Eq.~(\ref{eq:TTSH}). The contributions of the first two terms dominate over those of the third term when $\mathcal{P}_{Sk}>0$, as we show in detail in the Supplemental Material~\ref{SM}.

	\begin{figure*}[!]
			\resizebox{\linewidth}{!}{
			\includegraphics[scale=1]{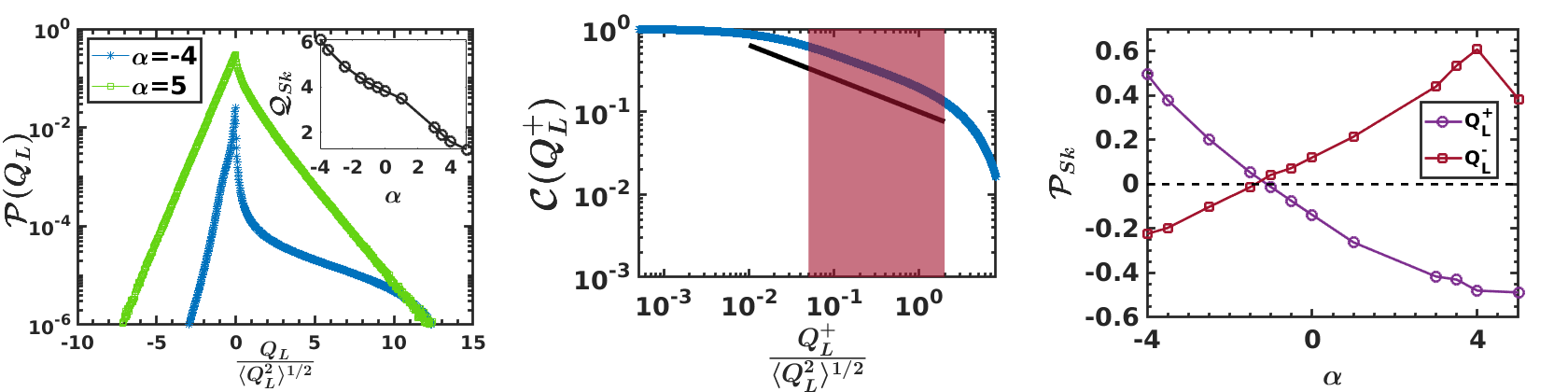}
		\put(-380,100){\rm {\bf(a)}}
		\put(-240,100){\rm {\bf(b)}}
		\put(-110,100){\rm {\bf(c)}}
		}
		\caption{(Color online) (a) Semi-log plots of $\mathcal{P}(Q_{L})$ for runs A1 (blue) and A13 (green). Inset gives the plot versus $\alpha$ of skewness, $\mathcal{Q}_{Sk}$, for $\mathcal{P}(Q_{L})$. (b) Log-log plot of $\mathcal{C}(Q^{+}_{L})$ for run A1; the shaded region shows a power-law and the solid black line gives the fit $\mathcal{C}(Q^{+}_{L})\sim [Q^+_L]^{-\vartheta}$, with $\vartheta = 0.37\pm0.04$. (c) Plots versus $\alpha$ of $\mathcal{P}_{Sk}$ for the conditioned PDFs (see text) $\mathcal{P}(p_{L}\vert Q^{+}_{L})$ (violet) and $\mathcal{P}(p_{L}\vert Q^{-}_{L})$ (maroon).}
		\label{fig:ow}
	\end{figure*}
	In 2D incompressible flows, the Okubo-Weiss parameter~\cite{okubo1970horizontal,weiss1991dynamics,giomi2015geometry,perlekar2011persistence} distinguishes between vortical and strain-dominated regions. We define this, along particle trajectories, as follows:
	\begin{equation}
		Q_{L}(t)=\frac{\omega^{2}-\sigma^{2}}{4}\bigg\vert_{x_{L}(t)},
	\end{equation}
	where $\omega^{2}=\frac{1}{2}\sum_{i,j}(\partial_{i}u_{j}-\partial_{j}u_{i})^{2}$ and $\sigma^{2}=\frac{1}{2}\sum_{i,j}(\partial_{i}u_{j}+\partial_{j}u_{i})^{2}$, with $i,j=1,2$.
	 $Q_{L}>0 ~(\equiv Q^{+}_{L})$, in vortical regions, and $Q_{L}<0 ~(\equiv Q^{-}_{L})$, in strain-dominated regions. The PDF $\mathcal{P}\big(\frac{Q_{L}}{\langle Q^{2}_{L}\rangle^{1/2}}\big)$, is positively skewed; its skewness $\mathcal{Q}_{sk}$ decreases with increasing $\alpha$, but remains positive throughout the range of $\alpha$ for Runs A1-A12 (inset of Fig.~\ref{fig:ow} (a)). For high activities, the cumulative PDF $\mathcal{C}(Q^{+}_{L})$, shows a power-law tail for $Q^{+}_{L}$ (Fig.~\ref{fig:ow} (b)), a unique feature of the bacterial turbulence we study~\footnote{Similar PDFs have been obtained in Ref.~\cite{singh2021lagrangian}, but the 
	 power-law form has not been noted.}; in contrast, for the high-friction regime ($\alpha>2$), the tail of $\mathcal{P}(Q_{L})$ has a faster-than-exponential decay, for small and positive values of $Q_{L}$, as in 2D fluid turbulence. Furthermore, in the large-activity regime $\alpha\leq-2$, the positivity of $\mathcal{P}_{Sk}$  arises from vortical regions, whereas, in the high-friction regime $\alpha\geq2$, this positive skewness comes from the strain-dominated regions, which we surmise from Fig.~\ref{fig:ow}(c), where we plot $ \mathcal{P}_{Sk}$ for the conditioned PDFs $\mathcal{P}(p_{L}\vert Q^{+}_{L})$ and $\mathcal{P}(p_{L}\vert Q^{-}_{L})$.
\\
		Quasi-2D experiments on dense suspension of aerobic bacteria, e.g., \textit{B. subtilis}, show that the average speed of  bacterial flow increases with the oxygen concentration~\cite{cisneros2011transition,sokolov2007,sokolov2012physical}. We can increase the activity by making $\alpha$ large and negative; in experiments, the activity can be increased by enhancing the oxygen, because the polar-ordered velocity scale $v_{p}=\sqrt{\frac{\vert\alpha\vert}{\beta}}$ is a measure of the swimming speed of bacteria; $u_{rms} \propto \alpha$ (cf. \cite{joy2020friction}); and in the frictional or $\alpha > 0$ regime, the value of $\alpha$ can be tuned in experiments by changing the bottom friction or the air-drag-induced friction (see the Supplemental Material~\ref{SM} for details). Therefore, experiments on dense bacterial suspensions should be able to examine irreversibility in bacterial turbulence as a function of the activity as we have done in Fig.~\ref{fig:Pskn_pdf}. 
		
		It is important to use the methods we describe here to explore irreversibility of bacterial turbulence in other models~\cite{linkmann2019phase,rana2020coarsening,oza2016generalized} and also in models for active fluids~\cite{chatterjee2021inertia,bowick2021symmetry} and active nematics~\cite{thampi2013velocity,thampi2016active,alert2020universal}. We propose to carry out such studies in the near future. 
		
	\newpage
	\section*{Supplemental Material}\label{SM}
	In this Supplementary Material we provide details of the following:
\begin{enumerate}
    \item Our direct numerical simulations (DNSs).
    \item Different contributions to the energy spectrum and the role of the advective term.
    \item The effects of filtering on the statistics of the power $p_L$.
    \item Additional figures and tables.
\end{enumerate}
	
\subsection*{Direct Numerical Simulation (DNS)} 
For two dimensional (2D) incompressible flows we rewrite Eq.~\ref{eq:TTSH}, in the main paper, in terms of the vorticity field $\bom(\textbf{x},t)$ as follows: 
\begin{align}
\frac{\partial{\bom}}{\partial t} +\lambda_{0}\bu\cdot\mathbf{\nabla}\bom&=-\alpha{\bom}-\beta\mathbf{\nabla}\times(|\bu|^{2}\bu) \nonumber \\
&+\Gamma_{0}\nabla^{2}{\bom}-\Gamma_{2}\nabla^{4}{\bom};
\label{eq:TTSH_vort}
\end{align}
and we define the stream function $\Psi(\textbf{x},t)$ in terms of which
\begin{equation}
\bu=\mathbf{\nabla}\times\Psi\hat{\textbf{z}} \;\; \rm{and} \;\; \bom=-\nabla^{2}\Psi ,
\label{eq:streamfn}
\end{equation}
where $\hat{\textbf{z}}$ is the unit vector perpendicular to the plane containing $\bu$.	
Our DNS of Eqs.~(\ref{eq:TTSH_vort}) and (\ref{eq:streamfn}) employs a pseudospectral method~\cite{canuto2012spectral,perlekar2011persistence}; we use a square simulation domain, with 
sides of length $L=2\pi$, periodic boundary conditions in all directions, and $N\times N$
 collocation points distributed uniformly over this domain. In most of our simulations we use $N=1024$; we have checked in representative simulations that our results are not altered significantly if  we use $N=2048$. We employ the second-order Runge-Kutta integrating-factor method IFRK2 for time marching~\cite{cox2002exponential}. We have developed a CUDA Fortran code for our DNSs which are excuted on K20 and K80 GPUs. 
  Once our DNS yields a turbulent, but statistically	steady, state, we introduce $N_p$ Lagrangian particles and follow their trajectories. The position $\mathbf{x}_{L}(t)$ of such a particle evolves as follows:
 	\begin{equation}
 		\frac{d \mathbf{x}_{L}(t)}{dt} \equiv \mathbf{v}_{L}(t) = \bu(\mathbf{x}_{L}(t),t).\nonumber
 		\label{eq:Lagpart}
 	\end{equation}
 	We integrate Eq.~(\ref{eq:Lagpart}) by using the second-order Runge-Kutta scheme and bi-linear interpolation to obtain the particle velocity $\mathbf{v}_{L}(t)$ at off-grid points. 
 	We evolve each particle trajectory for $3\times10^{6}$ time steps; and we store $\mathbf{x}_{L}$, $\mathbf{v}_{L}$, the particle acceleration $\mathbf{a}_{L}$, and the power $p_{L}$ after every $50$ iterations.
 	
 	We follow Refs.~\cite{joy2020friction,Wensink14308} in restricting our model parameters to experimentally realizable regimes. The average velocities observed in experiments on \textit{B. subtilis} are $ \simeq 25~\mu m/s$, at normal oxygen concentrations; the typical viewing area is $400 \mu m\times 400 \mu m$; we map these to the constant velocity scale $v_{0}$ and the simulation box area $L\times L$, respectively. This gives us the scale factors of $25/v_{0}$ and $4\times 10^{-2}/L$ for mapping velocities and lengths, respectively, in our DNS to their experimental counterparts: specifically, $v_0$, $v_p$, and $u_{rms}$ are, respectively, $\simeq 25~\mu m/s$, $\simeq (25~\mu m/s \; \rm{to} \; ~70~\mu m/s)$ and $\simeq (4.5~ \mu m/s \; \rm{to} \; 65~\mu m/s)$; similarly, $\Lambda$, which sets the linear
     scale for vortical regions, $\simeq 25 \mu m$.
    \subsection*{Energy budget}
       For the shell-averaged energy spectrum 
   \begin{equation}
       \mathcal{E}(k)=\frac{1}{2} \sum_{k'=k-1/2}^{k'=k+1/2}\langle\widetilde{\bu}(\mathbf{k}^{'}).\widetilde{\bu}(-\mathbf{k}^{'})\rangle_{t}
   \end{equation}
   we have~\cite{bratanov2015new}
   \begin{eqnarray}{\label{eq:espec}}
   	\partial_{t}\mathcal{E}(k)&=&T^{a}(k)-T^{c}(k)-2\alpha \mathcal{E}(k) \nonumber \\ &-&2\Gamma_{0}k^{2}\mathcal{E}(k)-2\Gamma_{2}k^{4}\mathcal{E}(k),
   	\label{eq:enbudget}
   \end{eqnarray}
   where $T^{a}(k)$ and $T^{c}(k)$, the $k$-shell averaged contributions from the advective and cubic terms in Eq.~(\ref{eq:TTSH_vort}), respectively, are
   	\begin{align}{\label{eq:enonlin}}
   		T^{a}(k)&=-\lambda_{0}\sum_{k'=k-1/2}^{k'=k+1/2}\langle\widetilde{\bu}(-\mathbf{k}^{'}).\textbf{P}(\mathbf{k'}).\widetilde{(\bu.\mathbf{\nabla}\bu)}(\mathbf{k'})\rangle_{t} \nonumber \\ 
   	   T^{c}(k)&=\beta\sum_{k'=k-1/2}^{k'=k+1/2}\langle \widetilde{\bu}(-\mathbf{k}^{'}).\textbf{P}(\mathbf{k'})\widetilde{(\vert \bu\vert^{2}\bu)}(\mathbf{k'})\rangle_{t},
   	\end{align}
    with $P_{ij}(\mathbf{k})=\delta_{ij}-\frac{k_{i}k_{j}}{k^{2}}$ the transverse projector and $\langle \cdot \rangle_{t}$ the average over time $t$. The flux of energy arising from the advective term is 
    \begin{equation}
    \Pi(k)=-\sum_{k'=0}^{k'=k}T^{a}(k')   .   
    \end{equation}
  
   The effective viscosity
	\begin{equation}
		k^2 \nu_{eff}(k)=\big(\alpha+2\beta u^{2}_{rms}+\Gamma_{0} k^2 +\Gamma_{2}k^4\big)\nonumber
	\end{equation}
	can be used to rewrite Eq.~(\ref{eq:TTSH_vort}), in a form that resembles the Navier-Stokes, with the constant viscosity $\nu$ replaced by $\nu_{eff}(k)$. To obtain Eq.~(\ref{eq:effvis}), we use the approximation $T^{c}(k)\simeq-4\beta u^{2}_{rms}\mathcal{E}(k)$ suggested in Ref.~\cite{bratanov2015new}; here, $u_{rms}$ must be obtained from our calculation. 
   Clearly, the wave numbers $k$ at which energy is injected
	(dissipated) are those with $\nu_{eff}(k)<0 ~(>0)$. 
   \\
   \begin{figure*}[!]
    		\resizebox{\linewidth}{!}{
    		\centering 
    		\includegraphics[scale=0.3]{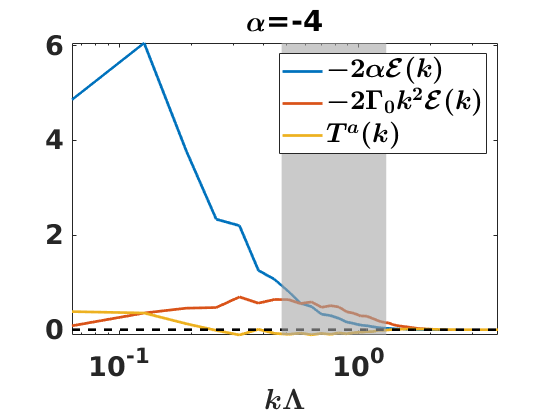}
    		\put(-110,90){\rm {\bf(a)}}
    		\includegraphics[scale=0.3]{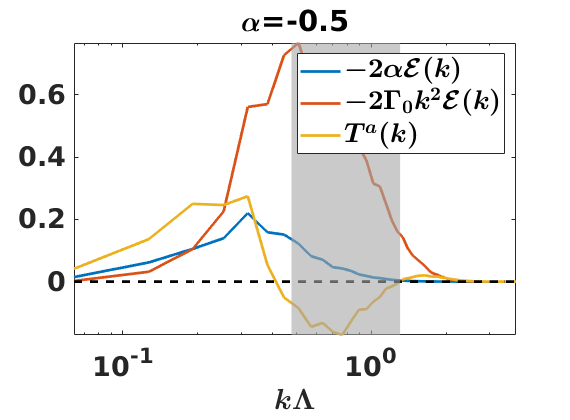}
    		\put(-110,90){\rm {\bf(b)}}
    		\includegraphics[scale=0.3]{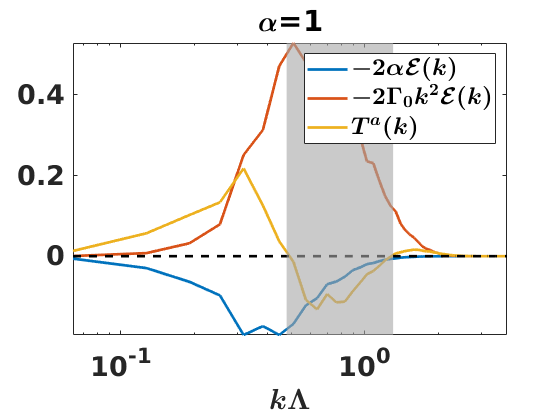}
    		\put(-110,90){\rm {\bf(c)}}
    		\includegraphics[scale=0.3]{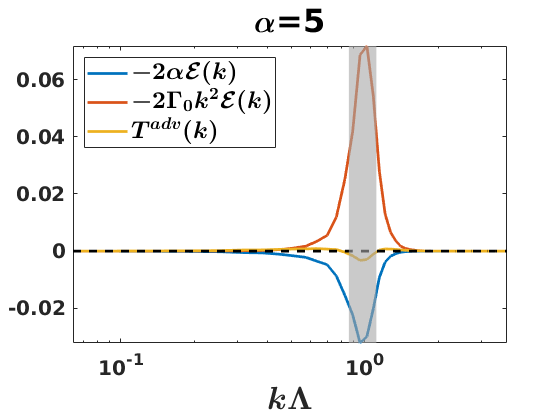}
    		\put(-110,90){\rm {\bf(d)}}
    	}
    	
    	\caption{(Color online) Semi-log plots versus $k\Lambda$ of $T^{a}(k)$ (yellow), $-2\alpha\mathcal{E}(k)$ (blue) and $-2\Gamma_{0} k^{2}\mathcal{E}(k)$ (orange); the gray-shaded areas indicate the ranges of $k$ for which $\nu_{eff} (k) < 0$ for (a) run A1, (b) run A7, (c) run A8 and (d) run A11}
    \end{figure*}\label{fig:ene_cont}
   	The sign of $\mathcal{P}_{Sk}$ (and, for small $\tau/\theta,$ the sign of $\mathcal{W}_{Sk}(\tau)$) displays the following correlation with the scale-by-scale energy budget in Fourier space, where we can identify the $k$-dependence of the energy contributions from the terms with coefficients $\alpha$, $\Gamma_0$, and $T^{a}(k)$ from Eq.~(\ref{eq:espec}) and (\ref{eq:enonlin}), which we show in Fig. (3): the contribution to the energy budget~(\ref{eq:enbudget}) from the active term, $-2\alpha\mathcal{E}(k)$, is significantly greater than $T^{a}(k)$, for values of $\alpha<-2$. For values of $\alpha>2$, where $\mathcal{P}_{Sk}>0$, the other active term, $-2\Gamma_{0}k^{2}\mathcal{E}(k)$ dominates over $T^{a}(k)$.
  
    \subsection*{Averaging and Filtering}
\begin{figure*}[!]	

	\resizebox{\linewidth}{!}{
    		\centering 
	\includegraphics[scale=0.4]{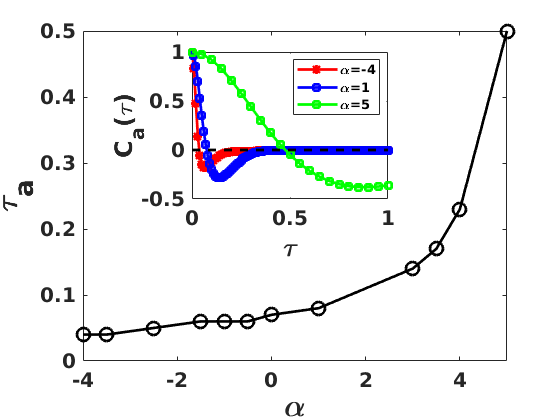}
	\put(-135,120){\rm {\bf(a)}}
	\includegraphics[scale=0.4]{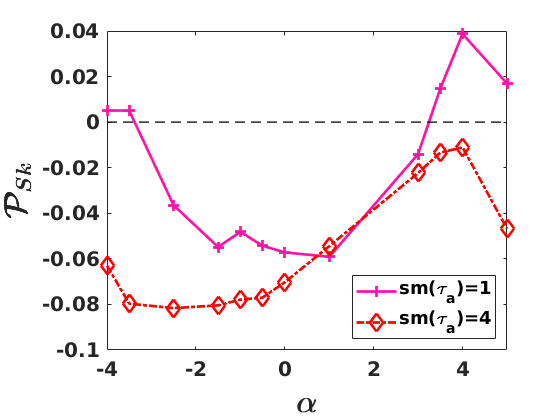}
	\put(-135,120){\rm {\bf(b)}}
	}
	\caption{(Color online) (a) Plot versus $\alpha$ of the acceleration-autocorrelation time $\tau_{a}$; the inset shows $C_{a}(\tau)$ as a function of $\tau$, for $\alpha=-4$ (blue), $\alpha=1$ (red), and $\alpha=5$ (green). (b) Plots versus $\alpha$ of $\mathcal{P}_{Sk}$ for $sm(\tau_{a})=1$ (solid line) and $sm(\tau_{a})=4$ (dashed line).}
	\label{fig:corr_acc}
\end{figure*}

\begin{figure*}[!]
    \centering
    \resizebox{\linewidth}{!}{
    {\includegraphics[scale=0.4]{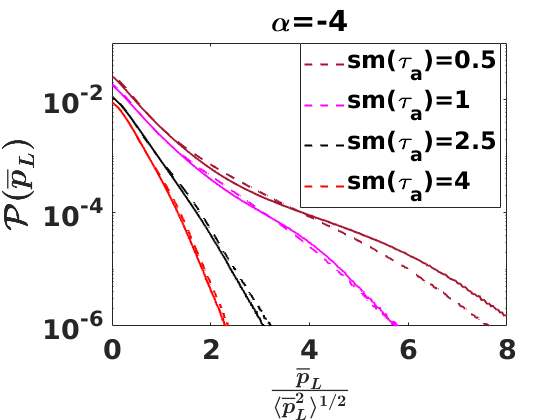}
    	\put(-135,120){\rm {\bf(a)}}

    }
    {\includegraphics[scale=0.4]{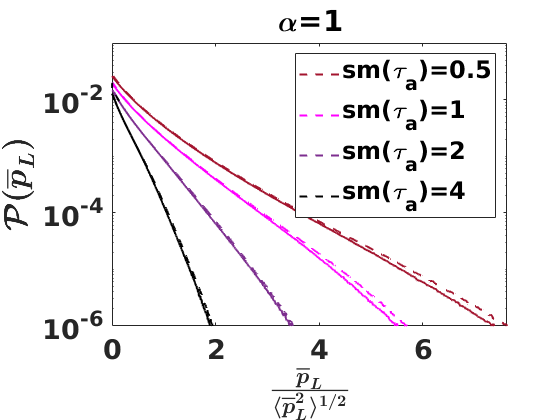}}
    	\put(-135,120){\rm {\bf(b)}}

    }
    
    \caption{Semi-log plots of the PDFs $\mathcal{P}(\overline{p}_{L})$ for different values $sm(\tau_{a})$ for (a) Run A1 and (b) Run A8.}
    \label{fig:smooth_pdf}
\end{figure*}

For conventional 2D fluid turbulence, the effects of different types of forcing, in both DNSs and experiments, on $\mathcal{P}_{Sk}$ have been discussed in Refs.~\cite{xu2014flight,pumir2014redistribution}, where it is noted that Faraday-wave experiments yield $\mathcal{P}_{Sk}>0$; this sign is attributed to the temporal coherence of Faraday waves. 
It is shown in Ref.~\cite{xu2014flight} that, if $p_L$ is averaged over a time that is comparable to this coherence time, then $\mathcal{P}_{Sk}<0$; this averaging filters high-frequency components in $p_L(t)$. Specifically, they use

\begin{equation}
	\overline{p}_{L}(t)=\frac{1}{\mathcal{T}}\int_{0}^{\mathcal{T}}p_{L}(t+t')dt';
\label{eq:pLaverage}
\end{equation}  
$\mathcal{T}$, the time over which $p_{L}(t)$ is averaged, is taken to be a multiple (typically $0.5 - 5$)  of the 
forcing-correlation time. In the mean-bacterial-velocity model, which we consider, there is no external forcing; the natural counterpart of $\mathcal{T}$ in Eq.~(\ref{eq:pLaverage}) is
the particle-acceleration time $\tau_a$ that we can obtain from the first zero-crossing of the normalised autocorrelation function
\begin{equation}
	C_{a}(\tau)=\frac{\langle a_{L}(t) a_{L}(t+\tau)\rangle}{\langle a_{L}(t) a_{L}(t)\rangle},
\label{eq:corr}
\end{equation}
which we plot in the inset of Fig.~\ref{fig:corr_acc} (a)for different values of $\alpha$.
In Fig.~\ref{fig:corr_acc} (a) we show that $\tau_{a}$ increase monotonically with $\alpha$. We define the smoothing parameter
\begin{equation}
sm(\tau_{a})\equiv\frac{\mathcal{T}}{\tau_{a}},
\label{eq:sm}
\end{equation}
the multiple of $\tau_{a}$ over which we average $p_{L}(t)$. In Fig.~\ref{fig:corr_acc} (b) we plot $\mathcal{P}_{Sk}$ versus $\alpha$ for $sm(\tau_{a})=1$ (solid line) and $sm(\tau_{a})=4$ (dashed line); for values of $\mathcal{P}_{Sk}>0$ (in both high-activity and high-friction regimes), the averaging defined in Eq.~\ref{eq:pLaverage} leads to a change in the sign of $\mathcal{P}_{Sk}$ for values of $sm(\tau_{a})>1$. This averaging does not change the sign of $\mathcal{P}_{Sk}$ if the unaveraged $p_L(t)$ leads $\mathcal{P}_{Sk}<0$. This is also evident from the PDFs of the filtered $\overline{p}_L$; we present representative plots for the PDFs for runs A1 and A8 in Figs.~\ref{fig:smooth_pdf} (a) and (b), respectively. From  Figs.~\ref{fig:corr_acc} and 
\ref{fig:smooth_pdf} we conclude that that the fast-gain-slow-loss events, shown in Fig. \ref{fig:color} of the main text, are filtered out by the averaging procedure, which we have described above, because these events occur over time scales that are $\simeq \tau_a$.


\subsection*{Supplemental Tables and Figures}
\begin{figure*}[!]	
    \resizebox{\linewidth}{!}{
    	\includegraphics[scale=0.4]{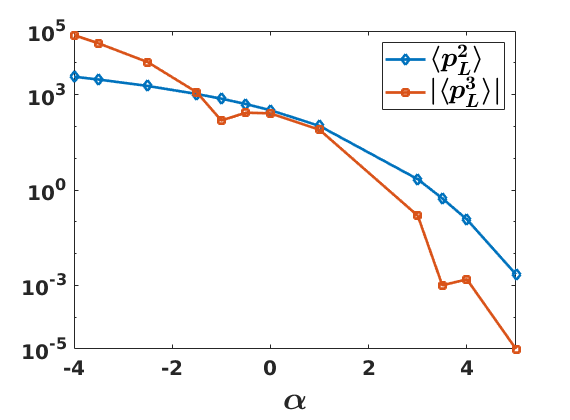}	
    		\put(-135,120){\rm {\bf(a)}}

    	\includegraphics[scale=0.4]{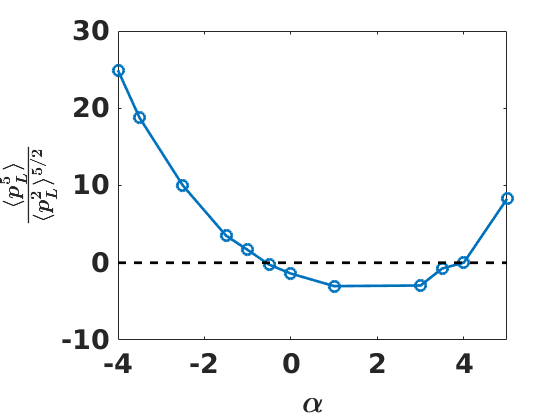}
    		\put(-135,120){\rm {\bf(b)}}

    	}
    \caption{ For Runs A1-A12 (Table I in the main paper): (a) Absolute values of the third moment $|\langle p^{3}_{L}\rangle|$ and the second moment $\langle p^{2}_{L}\rangle$ plotted versus $\alpha$. (b) Plots of the normalized fifth moment versus $\alpha$; these can also be used to quantify irreversibility and to draw the same conclusions as we have in the main paper. }
    	    	\label{fig:p3p2}
    	\end{figure*}

\begin{figure*}[!]
 	\resizebox{\linewidth}{!}{
    		\centering 
    		\includegraphics[scale=0.4]{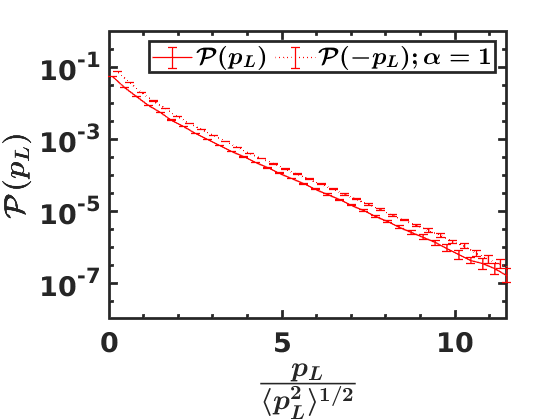}
    			\put(-135,120){\rm {\bf(a)}}

    		\includegraphics[scale=0.4]{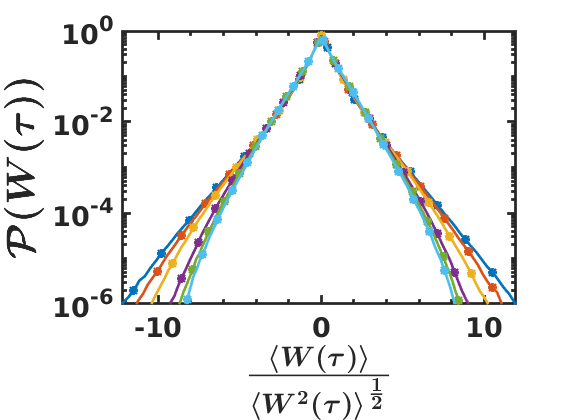}
    			\put(-135,120){\rm {\bf(b)}}

    		    		}
    	\caption{(Color online) Plots for Run C: $(a)$ Semi-log plot of the normalized PDF $\mathcal{P}(p_{L})$; negative values of $p_{L}$ (dashed) are reflected about the 
		vertical axis to highlight the asymmetry of $\mathcal{P}(p_{L})$. 
		$(b)$ Semi-log plots of $\mathcal{P}(W(\tau))$, for $\tau/\theta= 0.025, 0.08, 0.13, 0.25, 0.38$ and $0.50$ (outermost to innermost curve).
		 }
		 \label{fig:wpdf1}
    \end{figure*}

\begin{figure*}[!]
    	\resizebox{\linewidth}{!}{
    		\centering 
    		\includegraphics[scale=0.4]{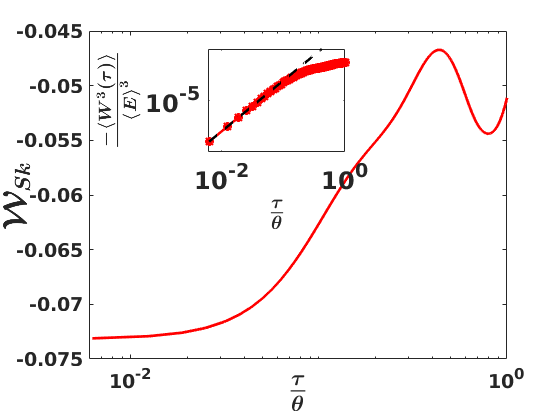}
    			\put(-135,130){\rm {\bf(a)}}

    		\includegraphics[scale=0.4]{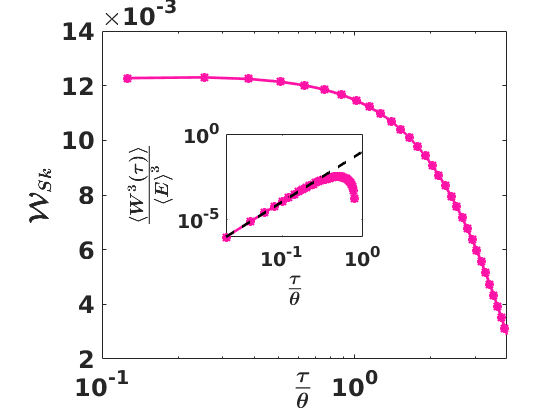}
    			\put(-135,130){\rm {\bf(b)}}

    		}
    	\caption{(Color online) Log-Log (base 10) plot versus $\tau/\theta$ of the skewness $\mathcal{W}_{Sk}(\tau)$. Inset: for the same range of $\tau/\theta$, a log-log plot versus $\tau/\theta$ of $\langle W^3(\tau)\rangle/\langle E \rangle^3$; the dashed black line is a fit to  $\langle W^3(\tau)\rangle/\langle E \rangle^3 \sim (\tau/\theta)^{3}$ for (a) Run C and (b) Run D.}
		\label{fig:ene_inc_si}

    \end{figure*}

\begin{figure}[!]
        \centering
        \resizebox{\linewidth}{!}{
        \includegraphics[scale=0.4]{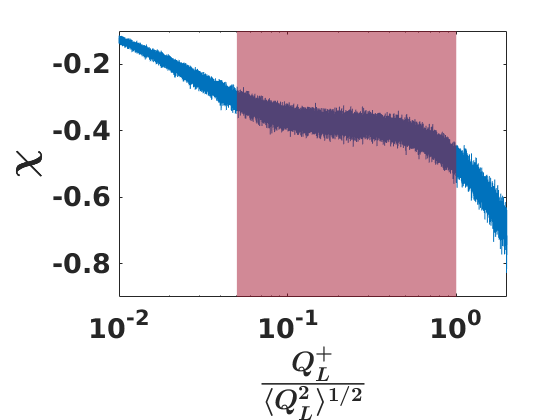}}
        \caption{For Run A1, semi-log plot of the local slope $\chi$ of $\mathcal{C}(Q^{+}_{L})$ in the shaded region that shows a power law.}
        \label{fig:ls_ow}
    \end{figure}

\begin{itemize}

\item 
In Table~\ref{tab:runD} we list the parameters for Run D in the main paper.

\item 
In Fig.~\ref{fig:p3p2} we plot various moments of $p_L$ versus $\alpha$ for Runs A1-A12.

\item
In Figs.~\ref{fig:wpdf1} and \ref{fig:ene_inc_si} we present representative plots for Runs C and D.

\item 
In Fig.~\ref{fig:ls_ow} we give a plot of the local slope in the power-law regime in $\mathcal{C}(Q^{+}_{L})$ for Run A1.
\end{itemize}
\begin{table}
		\centering
		\begin{tabular}{|c| c| c| c|c|c| c|}
			\hline
			Run & $u_{rms}$ & $L_{I}$& $T_{I}$& $\mathcal{P}_{Sk}$
			\\
		\hline	D&8.8e-2&4e-2&0.45&1.2e-2\\
			\hline
		\end{tabular}
		\caption{ DNS values of $u_{rms}$, $L_{I}$, $T_{I}$ and $\mathcal{P}_{Sk}$ for Run D.}
		\label{tab:runD}
	\end{table}

    

\subsection*{Acknowledgments}	
	
		We thank J.K. Alageshan, V. Deshpande (NVIDIA), N.B. Padhan, S. Shukla, and S.S.V. Kolluru for discussions, CSIR, the National Supercomputing Mission (NSM), and SERB (India) for support, and SERC (IISc) for computational resources. 

	\bibliography{references}
	
\end{document}